\documentclass[twocolumn,preprintnumbers,amsmath,amssymb,aps,prl,floatfix]{revtex4}
\usepackage{graphicx}
\usepackage{dcolumn}
\usepackage{amsmath}
\def\apj #1 #2 #3 {#1, ApJ, {\bf #2}, #3}
\def\apjl #1 #2 #3 {#1, ApJ, {\bf #2}, L#3}
\def\apjs #1 #2 #3 {#1, ApJS, {\bf #2}, #3}
\def\aap  #1 #2 #3 {#1, A\&A, {\bf #2}, #3}
\def\mnras #1 #2 #3 {#1, MNRAS, {\bf #2}, #3}
\def\pra #1 #2 #3 {#1, Phys.~Rev.~A., {\bf #2}, #3}
\def\prb #1 #2 #3 {#1, Phys.~Rev.~B., {\bf #2}, #3}
\def\prc #1 #2 #3 {#1, Phys.~Rev.~C., {\bf #2}, #3}
\def\prd #1 #2 #3 {#1, Phys.~Rev.~D., {\bf #2}, #3}
\def\pre #1 #2 #3 {#1, Phys.~Rev.~E., {\bf #2}, #3}
\def\prl #1 #2 #3 {#1, Phys.~Rev.~Lett., {\bf #2}, #3}
\def\plb #1 #2 #3 {#1, Phys.~Lett.~B., {\bf #2}, #3}
\def\science #1 #2 #3 {#1, Science., {\bf #2}, #3}
\def\nature #1 #2 #3 {#1, Nature., {\bf #2}, #3}
\def\nphysa #1 #2 #3 {#1, Nucl.~Phys.~A., {\bf #2}, #3}
\def\nphysb #1 #2 #3 {#1, Nucl.~Phys.~B., {\bf #2}, #3}
\def\nphysbs #1 #2 #3 {#1, Nucl.~Phys.~B.~Suppl., {\bf #2}, #3}

\def\h#1{\hbox{${}^{#1}$H}}

\def\h502{\hbox{$ h^{2}_{50}$}}

\def\la{\mathrel{\mathpalette\fun <}}
\def\ga{\mathrel{\mathpalette\fun >}}
\def\fun#1#2{\lower3.6pt\vbox{\baselineskip0pt\lineskip.9pt
  \ialign{$\mathsurround=0pt#1\hfil##\hfil$\crcr#2\crcr\sim\crcr}}}
\newcommand{\wbar}{{\overline w}}
\begin{document}
%\draft
%
\title{Stochastic Gravitational Wave Background in Brane World
Cosmology} 
\author{Kiyotomo Ichiki \footnote{E-mail address: ichiki@th.nao.ac.jp}}
\affiliation{%
Department of Astronomy, University of Tokyo, 7-3-1 Hongo, Bunkyo-ku,
Tokyo 113-0033, Japan}
\affiliation{%
Division of Theoretical Astrophysics, National Astronomical
Observatory, 2-21-1, Osawa, Mitaka, Tokyo 
181-8588, Japan}

\author{Kouji Nakamura   \footnote{E-mail address: kouchan@th.nao.ac.jp}}
\affiliation{%
Department of Astronomical Science, the Graduate University for
Advanced Studies, 2-21-1, Osawa, Mitaka, Tokyo 181-8588, Japan}
\date{\today}
\begin{abstract}%%%%%%%%%%%%%%%%%%%%%%%%%%%%%%%%%%%%%%%%%%%%%%%%%%%%%
 We investigate the cosmological evolution of gravitational waves in
 Friedman-Robertson-Walker brane world models embedded in a five
 dimensional anti de-Sitter spacetime. 
 To predict the spectrum of stochastic gravitational background at present
 generated in the inflationary phase, we numerically calculate the
 evolution of gravitational waves according to the method developed in
 our previous paper [K. Ichiki and K. Nakamura, arXiv:hep-th/0310282]. 
 The resulting logarithmic energy spectrum is significantly altered
 from that of the standard four dimensional inflationary models in the
 frequency range $f\ga f_{\rm AdS} \approx 10^{-4}$ Hz, which
 corresponds to the curvature scale we set of anti de-Sitter
 spacetime, $l \approx 1$mm. 
\end{abstract}
\maketitle

%%%%%%%%%%%%%%%%%%%%%%%%%%%%%%%%%%%%%%%%%%%%%%%%%%%%%%%%%%%%%%%%%
%\section{Introduction}
%\label{sec:intro}
In recent years much attention has been paid to interesting models of
spacetimes in which fields of the standard model are confined to a
``brane'', while only the gravitational field propagates in all the
dimensions called the ``bulk''. 
Since the proposal of a brane world model of our spacetime by Randall
and Sundrum~\cite{Randall:1999vf} (RS II model), the phenomenology of
brane world cosmological models has been the subject of intensive
investigations \cite{Binetruy:1999hy}.
In their models our four dimensional spacetime is embedded in the five
dimensional anti de-Sitter (AdS5) spacetime with curvature scale $l$,
and the four-dimensional gravity is effectively recovered in larger
scales than $l$. 
The current direct bound of the experimental gravitational probe allows
the scale of extra dimensions or curvature scale $l$ to be as large as
a (sub)millimeter \cite{Chiaverini:2002cb}.

% %********************************************************

In addition to the direct experimental probe of the Newton
potential, if our universe is really the RS II type brane universe,
there is a possibility that the evolution history of the universe is
significantly altered when the Hubble scale $H^{-1}$ is smaller than
$l$ in the early universe.
Actually, some authors have been discussed new physics in the
epoch of the early universe where Hubble scale is smaller than $l$ and
the possible sources of gravitational
waves\cite{Hogan:2000is,Inoue:2003di}.
In particular, it is pointed out that these new physics in the
early universe will appear in the stochastic gravitational wave
background with the characteristic frequency around $10^{-4}$
Hz\cite{Hogan:2000is}, if the scale of $l$ is order of $1$mm.

%********************************************************

Thus, the stochastic gravitational waves will be a promising candidate
which provides direct and deep probes of the
early universe (for example, see \cite{Maggiore:1999vm}).
One might see in principle earlier epochs of the universe than the
photon last scattering surface by gravitational waves.
In order to give theoretical predictions of the stochastic
gravitational waves, we have recently proposed the single null
coordinate system to solve the gravitational waves in the expanding
brane universe\cite{Ichiki:2003hf}, based on the idea in Ref.\cite{Nakamura:2002bz}.
We have concentrated on the brane world model of the flat
Friedmann-Robertson-Walker (FRW) universe without ``dark radiation''
following discussions in Ref.\cite{Ichiki:2002eh} and the fact that
the flat FRW universe is supported by recent precise measurements of
the cosmic microwave background (CMB) anisotropies~\cite{Spergel:2003cb}.
Though some authors have investigated based on the Gaussian normal
coordinate system in the neighborhood of the brane~\cite{Hiramatsu:2003iz}, 
this coordinate system includes a coordinate singularity in 
the bulk, where the metric function vanishes \cite{Ishihara:2001qe},
and the treatment of wave equation near the singularity becomes
delicate.
However, these delicate problems are removed by the introduction of the
coordinate system proposed in Ref.\cite{Ichiki:2003hf}.

% %*********************************************************

In this letter, we investigate how the presence of extra dimension of
RS type leaves an imprint on the stochastic gravitational wave
background at present.
We solve the evolution of gravitational waves in the brane world model
and predict the present energy spectrum, following the previous
proposal\cite{Ichiki:2003hf}.
%\section{The Basic Equations}
In terms of the single null coordinate system proposed in
\cite{Ichiki:2003hf}, the metric on the AdS$_5$ is given by
\begin{equation}
ds^2 = -F(\tau,\wbar)du^2-2F(\tau,\wbar)dud\wbar +r^2(\tau,\wbar)d\Sigma_0^2~.
\label{nullmetric}
\end{equation}
Here the metric functions $F(\tau,\wbar)$ and $r(\tau,\wbar)$ are
given by
\begin{eqnarray}
F(\tau(u,\wbar),\wbar)&=&\frac{r^2/l^2}{\sqrt{r^2/l^2 +{\dot a}^2}+\dot a}~,\\
r(\tau(u,\wbar),\wbar)&=&\frac{a(\tau)}{2}
\times\left[(1-A)e^{\wbar/l}+(1+A)e^{-\wbar/l}\right]~, \nonumber 
\end{eqnarray}
respectively, where $A\equiv \sqrt{1+l^2 H^2}$. Here $a(\tau)$ is scale factor of the
FRW brane whose time evolution is
given by the generalized Friedmann equation \cite{Binetruy:1999hy}
\begin{equation}
  H^2 = \left(\frac{\dot{a}}{a}\right)^2
  =\frac{8 \pi G_{\rm N}}{3}\rho
%  -\frac{K}{a^2}
  +\frac{\Lambda_{4}}{3}
  +\frac{\kappa_{5}^4}{36}\rho^2.
  \label{Friedmann}
\end{equation}
In Eq.(\ref{Friedmann}), $\dot a = da/d\tau$, $\tau$ is the comoving time
on the FRW brane, $\rho$ is the energy density on the brane, 
$\Lambda_{4}=\kappa_{5}^4 \lambda^2 /12 + 3 \Lambda_{5}/4$ is the 
cosmological constant induced on the brane, $G_{\rm N}=\kappa_{5}^4
\lambda / 48 \pi$ and $\kappa_{5}$ are the four-dimensional and
five-dimensional gravitational constants, respectively. 
We also assume $Z_2$ symmetry across the brane following to the original
RS II model. Hereafter we set $l=1$ mm for the purpose of numerical
calculations.

%********************************************************

Apart from the polarization tensor, the equation for gravitational waves
in the bulk is simply given by that for the five dimensional massless
scalar field, $\Box_5 h = 0$. 
As derived in Ref.\cite{Ichiki:2003hf}, this equation is equivalent to:
\begin{eqnarray}
 \partial_{u}h&=&r^{-3/2} \left[g
 \partial_{\wbar}h\right]^\wbar_0+r^{-3/2} \\ &\times&
 \left\{\int_0^\wbar \left(\partial_\wbar h \left( J- \frac{\partial
 g}{\partial \wbar}\right)-g\frac{k^2}{r^2} h \right) dw +
 C(u)\right\}~, \nonumber
\end{eqnarray}
where $-k^2$ is the eigen value of the Laplacian of $\Sigma_0$, and
\begin{eqnarray}
 g&:=&\frac{F}{2}r^{3/2}~, \\
 J&:=&\left(\frac{1}{2}\frac{\partial F}{\partial \overline w}
    +\frac{3F}{2}\frac{\partial \ln r}{\partial \overline w}
    -\frac{3}{2}\frac{\partial \ln r}{\partial u}\right)r^{3/2}~~.
\end{eqnarray}
$C(u)$ is determined by the boundary conditions of the Neumann type on
the brane at each time step
\begin{equation}
 C(u)=a^{3/2}F(\tau,w)\times \partial_\wbar h(u,\wbar)|_{\wbar=0}~~.
\label{eq:b.cond}
\end{equation}
Once given an initial spectrum on an initial null hypersurface in the
bulk, which is a $u=const$ hypersurface in the bulk,
Eqs.(\ref{Friedmann})-(\ref{eq:b.cond}) 
are enough to predict the evolution of gravitational
waves.

%********************************************************

As an initial condition of numerical examples, we follow the
inflationary scenario and we start numerical calculation by setting
$h=const.$ on a initial null hypersurface. 
This initial value is a solution to the wave equation in the long 
wavelength limit $k/aH \to 0$ \cite{Langlois:2000ns}. 
We should note, however, that, for $k\neq0$ mode, $h=const.$ is not 
an exact solution to the wave equation but an approximate one in the
situation $k/aH \ll 1$.
Further, as the universe expands and decelerates, the approximation
$k/aH \ll 1$ becomes violated.
Conversely, this also implies that $h=const.$ is an
appropriate 
initial condition in the long wavelength approximation, if we start
the calculation from the past much earlier time than that of the
horizon crossing of each mode on the brane.
For this reason, we start the calculation when the wave is far enough
outside the horizon with the initial condition $h=const.$ to keep the
validity of the long wavelength approximation $k/aH \to 0$.

%********************************************************

\begin{figure}
  \rotatebox{0}{\includegraphics[width=0.35\textwidth]{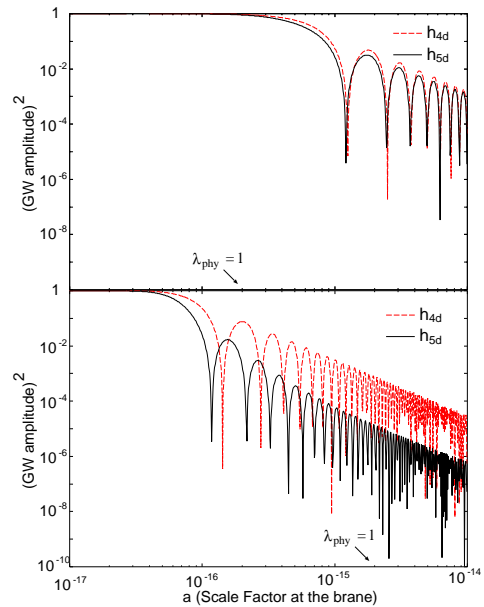}}
  \caption{The time evolution of gravitational waves
    with comoving frequency, $f=5\times 10^{-5}$[Hz] (upper panel) and 
    $f=5\times 10^{-4}$ [Hz] (lower panel). 
    The time evolution of gravitational wave with $f=5\times
  10^{-5}$[Hz]
    is almost ``four dimensional'' one, in which amplitude of
    gravitational waves scales as $h_{\rm 4D} \propto a^{-1}$.
    On the other hand, the amplitude of gravitational wave with
    $f=5\times 10^{-4}$ [Hz] in brane world model decreases
    faster than $a^{-1}$ when $\lambda_{\rm phy}\stackrel{<\,}{\sim}l$.
    After that, its evolution asymptotically becomes similar to that of
    standard model when $\lambda_{\rm phy}\stackrel{>\,}{\sim} l$.
  }
  \label{fig1}
\end{figure}

%\section{Physical Scales in the Problem and Numerical Results}
We show the examples of the evolution of gravitational waves $h_{\rm 5d}$ in
Fig.\ref{fig1}, which are obtained by solving
Eqs.(\ref{Friedmann})-(\ref{eq:b.cond}), numerically.
In these figures we also show 
``four dimensional gravitational waves'' $h_{\rm 4d}$ for comparison,
which is a solution of the four dimensional equation $\ddot h_{\rm 4d} +3 H
\dot h_{\rm 4d} +\left(k^2/a^2\right)h_{\rm 4d}=0$ and simply evolves as $a^{-1}$.

The feature of the evolutions of gravitational waves,
which one can easily see from Fig.\ref{fig1}, is that the amplitudes
of them are 
constant in time at first, and then start damping with oscillation. In
the standard four-dimensional cosmology, it is well known that the amplitude of
gravitational waves is ``frozen'' when the modes are  outside the
horizon ($k/aH \ll 1$), and then starts to oscillate after the modes enter the
horizon ($k/aH \gg 1$). 
As shown in Fig.\ref{fig1}, it turns out that this behavior is also
seen in the evolution of gravitational waves in brane world model
considered here.

% %****************************************************************

The other important feature is that the amplitude of five-dimensional
gravitational waves $h_{\rm 5d}$ decreases faster than that of standard
four-dimensional gravitational waves $h_{\rm 4d}$ in higher frequency
modes (the lower panel in Fig.\ref{fig1}). This is an essential feature of the
five-dimensional evolution of 
gravitational waves. As we mentioned above, the
gravitational waves with wavelength smaller than the curvature scale of
AdS$_5$ ($\lambda_{\rm phy} \equiv 2\pi a/k \la l$) would {\it feel} the
universe to be 
five dimensional, and  thus the evolution of gravitational waves with
such wavelength can be significantly altered. 
Even though the decomposition
of independent Kaluza-Klein (KK) modes is quite non-trivial in FRW brane
universe,  the strongly decelerating brane motion and Doppler effects by
moving brane cause the mode mixing
of gravitational waves, and do generate the effective ``KK modes''
which propagate into the bulk \cite{Hiramatsu:2003iz}. This is the
reason for the damping in amplitude of five-dimensional gravitational
waves relative to that of four-dimensional gravitational waves. %

% %*************************************************

\begin{figure}
\rotatebox{-90}{
\includegraphics[width=0.25\textwidth]{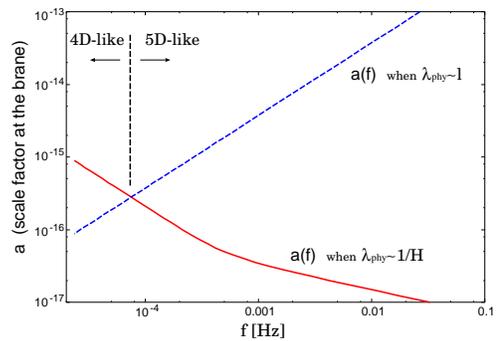}}
\caption{Cosmic scale factors at the brane as a function of comoving
 frequency when each Fourier mode
 re-enter the Hubble horizon ($\lambda_{\rm phy} = 1/H$; line) and
 when the wavelength becomes longer than the AdS5 curvature scale
 ($\lambda_{\rm phy} = l$; dashed line).}
\label{fig3}
\end{figure}

In Fig.\ref{fig3}, we show both the important criterions discussed above.
We depict the scale factors when each Fourier mode
with comoving frequency $f(=k/2\pi)$ enters the Hubble horizon ($\lambda_{\rm
phy} = 1/H$; line), and when their wavelength  becomes longer than
the curvature scale of AdS$_5$ ($\lambda_{\rm phys} = l$; dashed line)
in Fig.\ref{fig3}. 
A break around $f \approx 5\times 10^{-4}$ Hz in the line
$\lambda_{\rm phy}=1/H$ corresponds to the frequency
of the mode which crosses the Hubble horizon at the transition from
$\rho^2$-term dominate era to the standard radiation dominate era
(see, Eq.~(\ref{Friedmann})). 
These two lines cross each other ($\lambda_{\rm phy}=1/H=l$) at the
comoving frequency $f_{\rm AdS} \approx 10^{-4}$ Hz.
Figure \ref{fig3} is helpful to understand that gravitational
waves with comoving frequency larger than $f_{\rm AdS}$
Hz are affected by the existence of the fifth dimension, and the
gravitational waves change their behavior at $f=f_{\rm AdS}$

To explain this, let us first consider the lower
frequency modes, $f\la f_{\rm AdS}$. 
Figure \ref{fig3} shows that  the physical wavelength
$\lambda_{\rm phy}$ is larger than $l$ first, but the
mode is still far outside the horizon. 
After that, as the universe expands, the mode
crosses the horizon and starts to oscillate. 
Since the physical
wavelength is long enough relative to the curvature length of
AdS$_5$ at that time, the evolution of gravitational
waves is effectively four-dimensional one (see the upper panel in
Fig.\ref{fig1}).

On the other hand, the gravitational waves with comoving frequency
greater than $f_{\rm AdS}$ enter the Hubble horizon, first.
Then, they start to oscillate and their amplitude decreases
faster than $a^{-1}$ 
due to the ``KK mode'' mixing discussed above.
After that, as the universe expands the wavelength
of gravitational waves finally becomes longer than the curvature scale $l$
of AdS$_5$, and the time evolution of gravitational 
waves asymptotically mimics that of the standard four-dimensional FRW model,
i.e., the amplitude evolves as $h_{\rm 5d} \propto a^{-1}$. 
The evolution of gravitational waves in the lower panel of
Fig.{\ref{fig1}} does show this 
behavior.

\begin{figure}
\rotatebox{-90}{\includegraphics[width=0.25\textwidth]{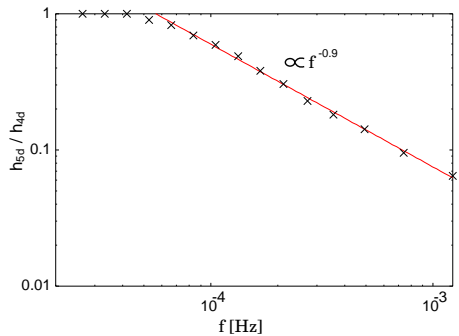}}
\caption{The amplitude ratio of gravitational waves ($h_{\rm 5d}$) to
 the standard four dimensional amplitude ($h_{\rm 4d} \propto
 a^{-1}$) determined in the asymptotic era ($\lambda_{\rm phy} \gg l$). Damping due to the momentum into the extra dimension becomes
 larger with larger comoving frequency.}
\label{fig4}
\end{figure}

In order to predict the spectrum of stochastic gravitational
waves at present, we have to know how much amplitude of gravitational waves in five
dimensional model has escaped from the brane  compared to that in four
dimensional model.
To accomplish this, we draw two envelope curves of the evolution of
gravitational waves for $h_{\rm 5d}$ and $h_{\rm 4d}$, and numerically
observe the ratio $h_{\rm 5d}/h_{\rm 4d}$ after the evolution of
gravitational waves in five dimensional model exhibits  asymptotic
behavior $h \propto a^{-1}$.
The result is depicted in Fig.\ref{fig4}.
Although there seem to exist some small
scatters, we find that the ratio can be well fitted
by a simple power law above some critical frequency:
\begin{equation}
 \frac{h_{\rm 5d}}{h_{\rm 4d}}=
\left\{
 \begin{array}{cl}
  1 &\quad\mbox{for}\ f<f_{\rm AdS}\\
  \left(\frac{f}{f_{\rm AdS}}\right)^{-\alpha}&\quad\mbox{for}\ f>f_{\rm AdS}\\
\end{array}
\right.
\label{eq:factor}
\end{equation}
where we can obtain from Fig.\ref{fig4} that  $f_{\rm AdS} \approx 10^{-4}$
Hz and $\alpha$ is roughly about $0.9$.

%

%\section{Discussion}
The spectrum of stochastic gravitational waves can be obtained by
multiplying the expected amplitude of four dimensional gravitational
waves today, $h_{\rm 4d}(\tau_0)$,
by the factor $\left(h_{\rm 5d}/h_{\rm 4d}\right)$ in Eq.(\ref{eq:factor}).
Present spectrum of four-dimensional gravitational waves $h_{\rm
4d}(\tau_0)$ can be obtained as follows.
In high energy era when $\rho^{2}$ term dominates, Hubble
parameter scales as $H\propto \rho_{r} \propto a^{-4}$
(see Eq.(\ref{Friedmann})) leading to $1/aH \propto a^3$. 
Using the fact that the amplitude of gravitational waves is almost
constant until horizon crossing $k/aH \sim 1$ and decreases as $h_{\rm
4d}\propto a^{-1}$ after that, we obtain $h_{\rm 4d} \propto k^{-1/3}$
in high energy era.
The same arguments lead that $h_{\rm 4d} \propto k^{-1}$ in standard
radiation dominated era, and $h_{\rm 4d} \propto k^{-2}$ in matter
dominated era \cite{Maggiore:1999vm}.
By combining these considerations, we can predict the final spectrum of
the stochastic gravitational waves.

\begin{figure}
\rotatebox{-90}{\includegraphics[width=0.32\textwidth]{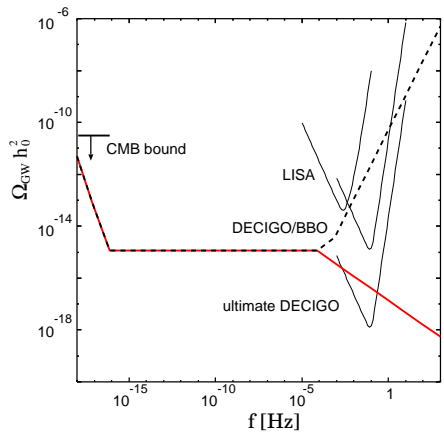}}
\caption{The spectrum of stochastic gravitational waves in brane world
 model with $l = 1$ mm (line) at present. The case in which we only
 include the effect of non-standard expansion law is shown by dashed
 line for comparison. Overall amplitude is normalized by CMB bound
 around the frequency $10^{-18}$ Hz. The noise curves of LISA,
 practical DECIGO/BBO  and quantum limited ultimate DECIGO for integration times
 $\approx 3$ yr are also shown in the figure \cite{Ungarelli:2000jp,LISA,Seto:2001qf,BBO}.}
\label{fig5}
\end{figure}

In Fig.\ref{fig5} we show the logarithmic energy spectrum of
gravitational waves, $\Omega_{\rm GW}(f) h_0^2$, where $\Omega_{\rm
GW}(f) \propto k^2 h^2$ is the energy density of gravitational waves per
logarithmic frequency interval in critical density units, and $h_0$ is
Hubble parameter today in units of $100$ km/s/Mpc.
We find that the spectrum becomes red, $\Omega_{\rm GW}h_0^2 \propto
f^{-0.46}$, in the frequency range  $f \ga f_{\rm AdS}$,
contrary to the expectation that the spectrum would have a spike due to
the non-standard (strongly decelerating) expansion law
\cite{Giovannini:1999qj}.

Finally we briefly mention the observational aspects. 
In standard models with CMB quadrupole anisotropies measured by COBE,
the energy 
density of gravitational waves is constrained to $\Omega_{\rm
GW} \la 10^{-15}$ at $f\sim 0.1$ Hz in standard cosmology. 
If future CMB experiments could directly measure tensor
amplitude and its spectral tilt ($n_T$) through temperature and polarization
anisotropies \cite{Eisenstein:1998hr},  one can in principle extrapolate and expect the amplitude of
gravitational waves to much higher frequency range, even up to $10^{15}$
Hz \cite{Turner:1996ck}. On the other hand, in the brane world models
considered in this paper, it is derived 
from our results that the spectral energy density will be
much smaller than expected above some critical frequency $f_{\rm AdS}
\approx  10^{-4}$ Hz. 
Unfortunately, stochastic radiation from white dwarf binaries prevents
us from detecting inflationary primordial gravitational waves in the
frequency range  $10^{-5} \la f \la 10^{-2}$ Hz. However, the sky would
be ``transparent'' to a primordial signal above the frequency $f \sim
0.1$ Hz by one year observation, if the neutron star - neutron star
merger rate is around $R \sim 10^{-5} $/yr/galaxy \cite{Ungarelli:2000jp}.
In this bands, the decihertz ($10^{-2} - 10$ Hz) interferometer
gravitational wave detectors, such as DECIGO \cite{Seto:2001qf} and BBO \cite{BBO}, have been proposed
aiming to detect the primordial gravitational waves as one of their
important goals.
Therefore, the signal of such brane world cosmological models might be
detected as a ``lack'' of gravitational waves in such higher 
frequencies. 

%{\it Acknowledgments.---}
K.I would like to thank M.~Yahiro and R.~Takahashi for helpful discussions. 
K.N. thanks M.~Omote and S.~Miyama for their continuous
encouragements. K.I.'s work is supported by Grant-in-Aid for JSPS Fellows.

%%%%%%%%%%%%%%%%%%%%%%%%%%%%%%%%%%%%%%%%%%%%%%%%%%%%%%%%%%%%%%%%%

%

\begin{thebibliography}{0}
%\cite{Randall:1999vf}
\bibitem{Randall:1999vf} 
L.~Randall and R.~Sundrum,
%``An alternative to compactification,''
Phys.\ Rev.\ Lett.\  {\bf 83}, 4690 (1999) 
%[arXiv:hep-th/9906064].
%%CITATION = HEP-TH 9906064;%%
%\cite{Mukohyama:1999wi}

%\cite{Binetruy:1999hy}
\bibitem{Binetruy:1999hy} 
P.~Binetruy, C.~Deffayet, U.~Ellwanger and D.~Langlois,
%``Brane cosmological evolution in a bulk with cosmological constant,''
Phys.\ Lett.\ B {\bf 477}, 285 (2000) 
%[arXiv:hep-th/9910219];
%%CITATION = HEP-TH 9910219;%%
%\cite{Kraus:1999it}
%\bibitem{Kraus:1999it}
P.~Kraus,
%``Dynamics of anti-de Sitter domain walls,''
JHEP {\bf 9912}, 011 (1999)
%[arXiv:hep-th/9910149];
%%CITATION = HEP-TH 9910149;%%
%\cite{Mukohyama:1999qx}
%\bibitem{Mukohyama:1999qx}
S.~Mukohyama,
%``Brane-world solutions, standard cosmology, and dark radiation,''
Phys.\ Lett.\ B {\bf 473}, 241 (2000)
%[arXiv:hep-th/9911165].
%%CITATION = HEP-TH 9911165;%%


%\cite{Chiaverini:2002cb}
\bibitem{Chiaverini:2002cb}
J.~Chiaverini, S.~J.~Smullin, A.~A.~Geraci, D.~M.~Weld and A.~Kapitulnik,
%``New experimental constraints on non-Newtonian forces below 100-mu-m,''
Phys.\ Rev.\ Lett.\  {\bf 90}, 151101 (2003)
%[arXiv:hep-ph/0209325].
%%CITATION = HEP-PH 0209325;%%

%\cite{Hogan:2000is}
\bibitem{Hogan:2000is}
C.~J.~Hogan,
%``Gravitational waves from mesoscopic dynamics of the extra dimensions,''
Phys.\ Rev.\ Lett.\  {\bf 85}, 2044 (2000)
%[arXiv:astro-ph/0005044].
%%CITATION = ASTRO-PH 0005044;%%

%\cite{Inoue:2003di}
\bibitem{Inoue:2003di}
K.~T.~Inoue and T.~Tanaka,
 %``Gravitational waves from sub-lunar mass primordial black hole binaries:  A
%new probe of extradimensions,''
Phys.\ Rev.\ Lett.\  {\bf 91}, 021101 (2003)
%[arXiv:gr-qc/0303058].
%%CITATION = GR-QC 0303058;%%


%\cite{Maggiore:1999vm}
\bibitem{Maggiore:1999vm}
M.~Maggiore,
%``Gravitational wave experiments and early universe cosmology,''
Phys.\ Rept.\  {\bf 331}, 283 (2000)
%[arXiv:gr-qc/9909001].
%%CITATION = GR-QC 9909001;%%


%\cite{Ichiki:2003hf}
\bibitem{Ichiki:2003hf}
K.~Ichiki and K.~Nakamura,
%``Causal structure and gravitational waves in brane world cosmology,''
arXiv:hep-th/0310282.
%%CITATION = HEP-TH 0310282;%%

% %\cite{Nakamura:2002bz}
\bibitem{Nakamura:2002bz}
K.~Nakamura,
%``Does a Nambu-Goto wall emit gravitational waves? Cylindrical Nambu-Goto  wall as an example of gravitating non-spherical walls,''
Phys.\ Rev.\ D {\bf 66}, 084005 (2002)
%[arXiv:gr-qc/0205031].
% %%CITATION = GR-QC 0205031;%%


%\cite{Ichiki:2002eh}
\bibitem{Ichiki:2002eh}
K.~Ichiki, M.~Yahiro, T.~Kajino, M.~Orito and G.~J.~Mathews,
%``Observational constraints on dark radiation in brane cosmology,''
Phys.\ Rev.\ D {\bf 66}, 043521 (2002)
%[arXiv:astro-ph/0203272].
%%CITATION = ASTRO-PH 0203272;%%

%\cite{Spergel:2003cb}
\bibitem{Spergel:2003cb}
D.~N.~Spergel {\it et al.},
 %``First Year Wilkinson Microwave Anisotropy Probe (WMAP) Observations:
%Determination of Cosmological Parameters,''
Astrophys.\ J.\ Suppl.\  {\bf 148}, 175 (2003)
%[arXiv:astro-ph/0302209].
%%CITATION = ASTRO-PH 0302209;%%


%\cite{Hiramatsu:2003iz}
\bibitem{Hiramatsu:2003iz}
T.~Hiramatsu, K.~Koyama and A.~Taruya,
 %``Evolution of gravitational waves from inflationary brane-world : Numerical
%study of high-energy effects,''
Phys.\ Lett.\ B {\bf 578}, 269 (2004)
%[arXiv:hep-th/0308072];
%%CITATION = HEP-TH 0308072;%%
%\cite{Battye:2003ks}
%\bibitem{Battye:2003ks}
R.~A.~Battye, C.~Van de Bruck and A.~Mennim,
%``Cosmological tensor perturbations in the Randall-Sundrum model:
%Evolution in
%the near-brane limit,''
Phys.\ Rev.\ D {\bf 69}, 064040 (2004)
%[arXiv:hep-th/0308134];
%\bibitem{Kobayashi:2003cn}
T.~Kobayashi, H.~Kudoh and T.~Tanaka,
%``Primordial gravitational waves in inflationary braneworld,''
Phys.\ Rev.\ D {\bf 68}, 044025 (2003)
%[arXiv:gr-qc/0305006];
%%CITATION = GR-QC 0305006;%%
%%CITATION = HEP-TH 0308134;%%
%\cite{Easther:2003re}
%\bibitem{Easther:2003re}
R.~Easther, D.~Langlois, R.~Maartens and D.~Wands,
%``Evolution of gravitational waves in Randall-Sundrum cosmology,''
JCAP {\bf 0310}, 014 (2003)
%[arXiv:hep-th/0308078].
%%CITATION = HEP-TH 0308078;%%
%

%\cite{Ishihara:2001qe}
\bibitem{Ishihara:2001qe}
H.~Ishihara,
%``Big bang of the brane universe,''
Phys.\ Rev.\ D {\bf 66}, 023513 (2002)
%[arXiv:gr-qc/0107085].
%%CITATION = GR-QC 0107085;%%





%\cite{Langlois:2000ns}
\bibitem{Langlois:2000ns}
D.~Langlois, R.~Maartens and D.~Wands,
%``Gravitational waves from inflation on the brane,''
Phys.\ Lett.\ B {\bf 489}, 259 (2000)
%[arXiv:hep-th/0006007].
%%CITATION = HEP-TH 0006007;%%


%\cite{Giovannini:1999qj}
\bibitem{Giovannini:1999qj}
M.~Giovannini,
%``Spikes in the relic graviton background from quintessential
        %inflation,''
Class.\ Quant.\ Grav.\  {\bf 16}, 2905 (1999)
%[arXiv:hep-ph/9903263].
%%CITATION = HEP-PH 9903263;%%

%\cite{Eisenstein:1998hr}
\bibitem{Eisenstein:1998hr}
D.~J.~Eisenstein, W.~Hu and M.~Tegmark,
 %``Cosmic Complementarity: Joint Parameter Estimation from CMB Experiments and
%Redshift Surveys,''
Astrophys.\ J.\  {\bf 518}, 2 (1998)
%[arXiv:astro-ph/9807130].
%%CITATION = ASTRO-PH 9807130;%%

%\cite{Turner:1996ck}
\bibitem{Turner:1996ck}
M.~S.~Turner,
%``Detectability of inflation-produced gravitational waves,''
Phys.\ Rev.\ D {\bf 55}, 435 (1997)
%[arXiv:astro-ph/9607066].
%%CITATION = ASTRO-PH 9607066;%%

%\cite{Ungarelli:2000jp}
\bibitem{Ungarelli:2000jp}
C.~Ungarelli and A.~Vecchio,
%``High energy physics and the very-early universe with LISA,''
Phys.\ Rev.\ D {\bf 63}, 064030 (2001)
%[arXiv:gr-qc/0003021].
%%CITATION = GR-QC 0003021;%%

%\cite{Seto:2001qf}
\bibitem{Seto:2001qf}
N.~Seto, S.~Kawamura and T.~Nakamura,
 %``Possibility of direct measurement of the acceleration of the universe  using
%0.1-Hz band laser interferometer gravitational wave antenna in  space,''
Phys.\ Rev.\ Lett.\  {\bf 87}, 221103 (2001)
%[arXiv:astro-ph/0108011].
%%CITATION = ASTRO-PH 0108011;%%

%\cite{BBO}
\bibitem{BBO}
http://universe.gsfc.nasa.gov/

%\cite{LISA}
\bibitem{LISA}
http://lisa.jpl.nasa.gov/index.html

\end{thebibliography}
\end{document}